\documentclass[ twocolumn,amssymb,aps,superscriptaddress,prl]{revtex4-1}

\usepackage{fullpage}
\usepackage{graphicx, subfigure}
\usepackage{epstopdf}
\usepackage{bm}
\usepackage{times}
\usepackage{amsfonts}
\usepackage{amssymb}
\usepackage{amsmath}
\usepackage{multirow} 
\usepackage{soul}
\usepackage{color}

\definecolor{LightGreen}{RGB}{144, 238, 144}
\sethlcolor{LightGreen}

\newcommand{\dd}{\text{d}}
\newcommand{\shat}{\hat{\mathbf{s}}} 
\newcommand{\phat}{\hat{\mathbf{p}}}

\newcommand{\zhat}{\hat{\mathbf{z}}}
\newcommand{\xhat}{\hat{\mathbf{x}}}
\newcommand{\yhat}{\hat{\mathbf{y}}}

\newcommand{\Mz}{M_{\left\{\zhat\right\}}}

\newcommand{\Etmpm}{\mathbf{E}_{m,p;\overline{-p}}^{\mathbf{t}}}
\newcommand{\Etmpp}{\mathbf{E}_{m,p;\overline{p}}^{\mathbf{t}}}
\newcommand{\Emp}{\mathbf{E}_{m,p}^{\mathbf{in}}}
\newcommand{\Emmmp}{\mathbf{E}_{-m,-p}^{\mathbf{in}}}
\newcommand{\Efmp}{\mathbf{E}_{m,p}^{\mathbf{in}\mathrm{,f}}}

\newcommand{\Etmp}{\mathbf{E}_{m,p}^{\mathbf{t}}}
\newcommand{\Etfmp}{\mathbf{E}_{m,p}^{\mathbf{t}\mathrm{,f}}}

\newcommand{\Sop}{\overline{\mathbf{S}}}
\newcommand{\Slens}{\overline{\mathbf{S}}_{\mathrm{lens}}}

\newcommand\sphat{\mathbf{\hat{\sigma}_p}}
\newcommand\smphat{\mathbf{\hat{\sigma}_{-p}}}
\newcommand\spphat{\mathbf{\hat{\sigma}_{+}}}
\newcommand\spmhat{\mathbf{\hat{\sigma}_{-}}}
\newcommand{\Imp}{I_{m,p;\overline{p}}^t}
\newcommand{\Impm}{I_{m,p;\overline{-p};}^t}
\newcommand{\gmp}{\gamma_{m,p}}

\begin{document}

\title{Far-field measurements of vortex beams interacting with nanoholes}

\author{Xavier Zambrana-Puyalto}
\email{xavislow@protonmail.com}
\affiliation{Department of Physics and Astronomy, Macquarie University, 2109 NSW, Australia }
\affiliation{ARC Centre for Engineered Quantum Systems, Macquarie University, 2109 NSW, Australia}
\affiliation{Aix-Marseille Universit\'{e}, CNRS, Centrale Marseille, Institut Fresnel UMR 7249, 13013 Marseille, France}
\author{Xavier Vidal}
\affiliation{Department of Physics and Astronomy, Macquarie University, 2109 NSW, Australia }
\author{Ivan Fernandez-Corbaton}
\affiliation{Institut of Nanotechnology, Karlsruhe Institute of Technology, 76021 Karlsruhe, Germany }
\author{Gabriel Molina-Terriza}
\email{gabriel.molina-terriza@mq.edu.au}
\affiliation{Department of Physics and Astronomy, Macquarie University, 2109 NSW, Australia }
\affiliation{ARC Centre for Engineered Quantum Systems, Macquarie University, 2109 NSW, Australia}

\begin{abstract}
We measure the far-field intensity of vortex beams going through nanoholes. The  process is analyzed in terms of helicity and total angular momentum. It is seen that the total angular momentum is preserved in the process, and helicity is not. We compute the ratio between the two transmitted helicity components, $\gmp$. We observe that this ratio is highly dependent on the helicity ($p$) and the angular momentum ($m$) of the incident vortex beam in consideration. Due to the mirror symmetry of the nanoholes, we are able to relate the transmission properties of vortex beams with a certain helicity and angular momentum, with the ones with opposite helicity and angular momentum. Interestingly, vortex beams enhance the $\gmp$ ratio as compared to those obtained by Gaussian beams.
\end{abstract}

\maketitle
The interaction of light with nano-apertures is a problem that has been carefully studied by many scientists. In particular, the work of Ebbesen \textit{et al.} \cite{Ebbesen1998} showing that nano-apertures could have an extraordinary transmission due to the coupling of light and surface plasmon polaritons (SPPs) opened up lots of possibilities among different fields in optics, micro-manipulation, biophysics and condensed-matter \cite{Mock2002,Mathieu2011,Mcdonnell2001,Haes2002}. Lots of these studies have focused on how the nano-apertures couple with SPPs depending on many different parameters \cite{Zayats2005,Genet2007}. Some others have focused on the radiation diagram of these nano-apertures and have studied how light with linear polarization normal ($\shat$) or parallel ($\phat$) to the plane of incidence are transmitted through the apertures both theoretically and numerically \cite{Aigouy2007,Lalanne2009,Ivan2011,Sol2012,Yi2012}. Finally, sub-wavelength nano-apertures have also been used to study the interaction between SPPs and the angular momentum (AM) of light \cite{Bliokh2008,Yuri2008,Vuong2010,Yuri2013,Sun2014,Zambrana2014Nat}.

Even though the first studies about the AM of light date back to the beginning of the twentieth century \cite{Poynting1909}, it was not until the 1990s when its use rapidly extended across different disciplines. The seminal finding that triggered much of the following developments was carried out by Allen and co-workers. In \cite{Allen1992}, the authors established a connection between the topological charge of paraxial vortex beams and their AM content. The finding implied that the AM content of optical beams could be controlled using available holography techniques - first Computer Generated Holograms (CGHs) and later Spatial Light Modulators (SLMs) \cite{Leseberg1987,Vasara1989,Bazhenov1990,Heckenberg1992OL,Dholakia2002,Grier2003}. Since then, the AM of light has been used in many diverse fields such as quantum optics \cite{Mair2001,Gabi2007}, optical manipulation \cite{He1995prl,Simpson1996}, optical communications \cite{Bo2007,Tamburini2012} or astrophysics \cite{Harwit2003,Tamburini2011}. Here, we present for the first time experimental far-field intensity recordings of the transmission of vortex beams through single nanoholes. We project the transmitted field into its two helicity components. Then, we observe that the mirror symmetry of the nanoholes constrains the transmission process of different modes up to a large extent. As it will be shown later on, the transmitted intensity of vortex beams with total AM $m$ and helicity $p$ is equal to the transmitted intensity of vortex beams with AM $-m$ and helicity $-p$. Finally, we compute the ratio between the two transmitted helicity components, which we denote as $\gmp$ and whose definition is given by eq. (\ref{gamma}). Not only we observe an enhanced helicity transference with respect to a Gaussian excitation \cite{Nora2014}, but also when investigating how this quantity changes with the size of the hole, we observe that the curves present structural differences.

The article is organised as follows. First, the optical set-up used to carry out the measurements is described. Second, the beams of light used to excite the nanoholes are mathematically characterized. Third, the interaction between the incident light and the sample is explained from the point of view of symmetries and conserved quantities. Then, a description of the methodology used to measure the transmitted light is given. Finally, the far-field measurements of the transmission of vortex beams through the nanoholes are presented and discussed.
\begin{figure}[htbp]
\centering\includegraphics[width=\columnwidth]{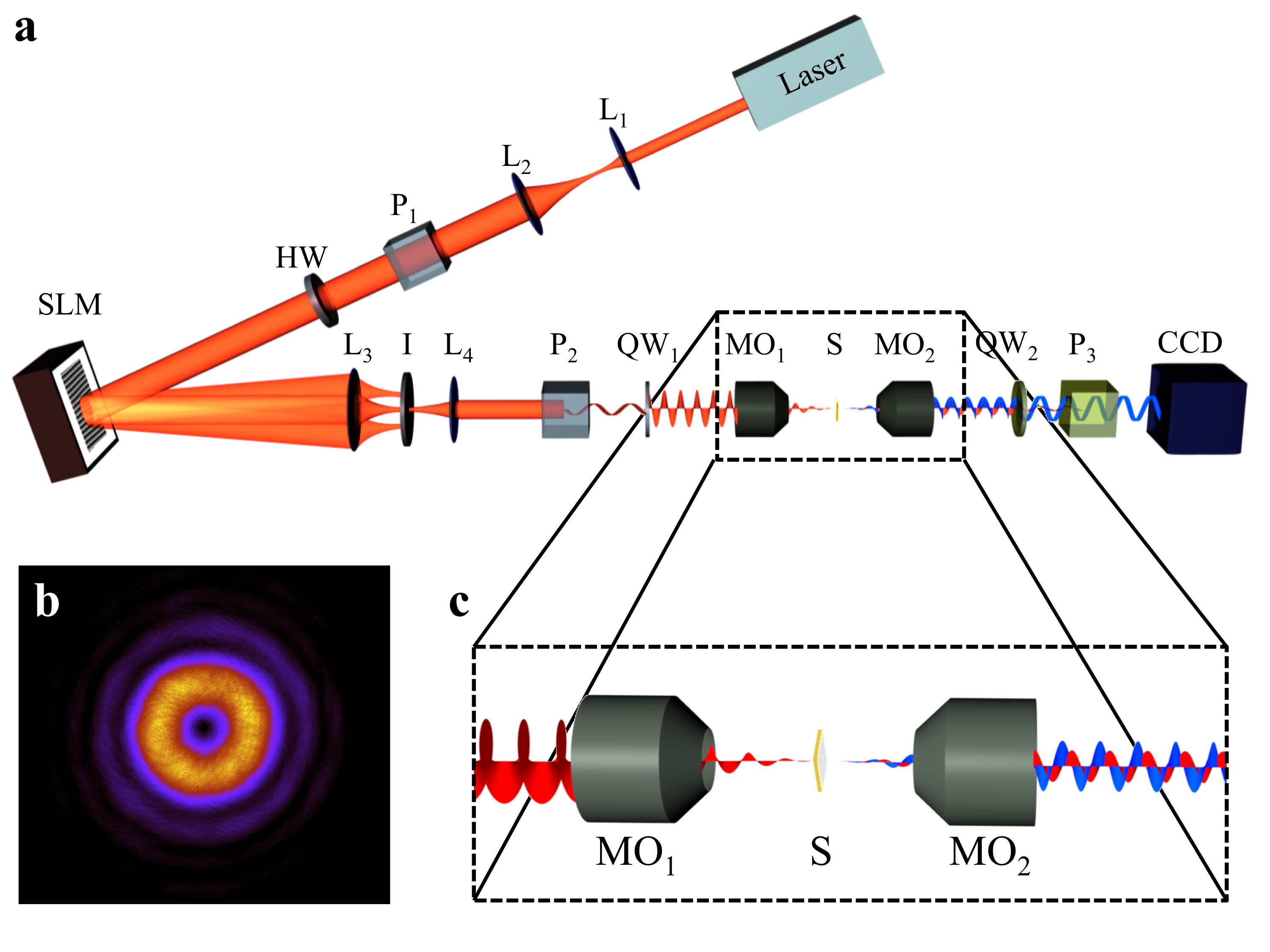}
\caption{(a) Schematic view of the optical set-up in consideration. A vortex beam with a well-defined AM and helicity goes through a sub-wavelength circular aperture and its transmissivity is recorded with a CCD camera. (b) Intensity profile of one of the vortex beams used in the experiment. (c) Schematics of the helicity change process. An incident paraxial beam has helicity $p$ (red) before and after MO$_1$. Then, it interacts with the sample S. The transmitted field has two helicity components, $p$ (red) and $-p$ (blue). Finally, the collimating objective MO$_2$ does not change the helicity content of the beam.}
\label{set-up}
\end{figure}

The experimental set-up is similar to the one used in \cite{Zambrana2014Nat}, and it is depicted in Figure \ref{set-up}. A CW laser operating at $\lambda=633$nm is used to generate a light beam. The laser produces a collimated, linearly polarized Gaussian beam. The beam is expanded with a telescope (lenses L$_1$-L$_2$) to match the dimensions of the chip of an SLM. After the telescope, the polarization state of the Gaussian beam is modified with a linear polarizer (P$_1$) and a half-wave plate (HW$_1$) to maximize the efficiency of the SLM. The SLM creates a vortex beam (see Figure \ref{set-up}b) by displaying an optimized pitchfork diffraction grating \cite{Bazhenov1990,Heckenberg1992OL,Leach2003,Zambrana2014Nat}. Proper control of the pitchfork hologram allows for the creation of a phase singularity of order $l$ in the center of the beam, \textit{i.e.} the phase of the beam twists around its center from $0$ to $2 \pi l$ radians in one revolution. Note that when $l=0$, the SLM behaves simply as a mirror. Because the SLM produces different diffraction orders, a modified 4-f filtering system with the lenses L$_3$-L$_4$ and an iris (I) in the middle is used to filter the non-desired orders of diffraction. Lens L$_3$ Fourier-transforms the beam, and the iris selects the first diffraction order and filters out the rest. Then, lens L$_4$ is used to match the size of the back-aperture of the microscope objective that will be used to focus the beam down to the sample. Before focusing, the preparation of the input beam is finished by setting its polarization to either left circular polarization (LCP) or right circular polarization (RCP). This is done with a linear polarizer (P$_2$) and a quarter-wave plate (QW$_1$). Since the beam is collimated, this change of polarization does not appreciably affect the spatial shape of the input beam. At this point, a water-immersion (NA$=1.1$) microscope objective (MO$_1$) is used to focus the circularly polarized paraxial vortex beam onto the sample (S). The sample is a set of single isolated nanoholes drilled on a 200nm gold film, deposited on top of a microscope slide. The sizes of the nanoholes range from 200-450nm (see Table \ref{Tgammas}), and they are located 50$\mu$m apart from their closest neighbours. Only one single nanohole is probed at a time, and the beam is centered with respect to the nanohole with a nanopositing stage. The transmitted light through the nanohole, which is scattered in all directions, is collected with MO$_2$, whose NA$=0.9$. MO$_2$ collimates the transmitted light, \textit{i.e.} the transmitted beam is paraxial once it leaves MO$_2$. Afterwards, the beam goes through another quarter-wave plate (QW$_2$), which transforms the two orthogonal circular polarization states into two orthogonal linear ones. A linear polarizer P$_3$ is placed after QW$_2$, which acts as an analyser. That is, P$_3$ selects one of the two circular polarization states existing before QW$_2$. Finally, a CCD camera records the intensity of the selected component. Notice that the main difference with respect to the set-up used in \cite{Zambrana2014Nat} is in the detection scheme, which allows for a complete characterization of the polarization of the collimated transmitted beam. \\

As shown in Fig.\ref{set-up}a, we prepare monochromatic paraxial circularly polarized vortex beams, which are tightly focused onto the nano-apertures. The mathematical expression of their electric field (before focusing) is given by:
\begin{equation}
\Emp \simeq A_{m,p}\rho^{\vert m-p \vert } e^{\left( i(m-p)\phi + i k z \right)} e^{(-\rho^2/w_0^2)} \sphat
\label{Epq}
\end{equation}
where $\sphat = (\xhat + i p \yhat) / \sqrt{2}$ is the circular polarization unitary vector,  with $\xhat, \yhat$ being the horizontal and vertical polarization vectors, and $p=\pm 1$ the handedness of the beam; $A_{m,p}$ is a normalisation constant; $w_0$ is the beam waist; $k$ is the wavenumber, $k=2 \pi / \lambda$ with $\lambda$ the single wavelength in consideration; and $(\rho$, $\phi$, $z)$ are the cylindrical coordinates. An implicit harmonic $\exp(-i \omega t)$ dependence is assumed, where $\omega = 2\pi c /\lambda$ is the angular frequency of light, and $c$ is the speed of light in vacuum. Notice that $m-p$ is the topological charge of the beam created by the SLM, \textit{i.e.} $m-p=l$. The origin of this formula can be seen in the collimated limit of Bessel beams with well defined helicity \cite{Ivan2012PRA,Nora2014}. Also, note that given the definition of $\sphat$, $p=1$ refers to LCP and $p=-1$ to RCP. In fact, it can be proven that $p$ is the value of the helicity of the beam in the paraxial approximation \cite{Ivan2012PRA,Nora2014}. In this approximation, $\Emp$ are eigenstates of the helicity operator. The helicity operator can be expressed as $\Lambda = (\nabla \times)/k $ for monochromatic beams \cite{Ivan2012PRA,Ivan2014}. Thus, $\Lambda \Emp  \approx p \Emp$ \cite{Ivan2012PRA,Nora2014}. This fact is very useful for our purposes, as it enables us to identify collimated circularly polarized vortex beams as states of well-defined helicity. States of well-defined helicity have eigenvalue $+1$ ($-1$) when their plane wave decomposition yields only left (right) handed polarized waves \cite{Ivan2012PRA}\cite[p170]{Tung1985}. Since the helicity operator is the generator of duality transformations \cite{Calkin1965,Cameron2012,Ivan2013}, $\Emp$ will be invariant under duality transformations within the paraxial approximation. In a similar manner, $\Emp$ are eigenstates of the $z$ component of the AM operator, $J_z$. That is, $J_z \Emp = m \Emp$. Then, because $J_z$ is the generator of rotations along the $z$ axis \cite{Rose1955,Tung1985}, $\Emp$ will be symmetric under these transformations, too. \\

In order to understand the forthcoming results, here we describe the light-matter interaction from the point of view of symmetries and conserved quantities \cite{Ivan2012PRA}. The main advantage of this description is that it easily allows for the prediction of different light-matter interaction in a qualitative way. In contrast, if quantitative predictions need to be made, full Maxwell equations (or dyadic Green tensor) solvers are needed. 

The nanoholes used in the experiment are almost perfectly round. Therefore, they are symmetric under rotations along the optical axis, which we will denote as $z$ without loss of generality. In addition, the sample is also symmetric under mirror transformations with respect to any plane containing the $z$ axis. 
Now, we will consider that all the properties of the sample are inherited by a linear integro-differential operator $\Sop$, which can be found using the Green Dyadic formalism. Then, $\Sop$ fulfils the following commutation rules:
\begin{equation}
\left[ \Sop, R_z \right]=\left[ \Sop, J_z \right]=\left[ \Sop, \Mz \right]=0
\label{S_comm}
\end{equation}
where $R_z$ and $\Mz$ are the operators that generate rotations along the $z$ axis and mirror transformations with respect to a plane that contains the $z$ axis, respectively. The first equality is a consequence of the fact that $J_z$ is the generator of rotations along the $z$ axis, \textit{i.e.} $R_z(\theta)=\exp(-i \theta J_z)$. Because of this, if a nanohole interacts with a beam which is an eigenstate of $R_z$, $J_z$, or $\Mz$ with eigenvalue $\nu$, the result of the interaction will still be an eigenstate of the same operator with the same eigenvalue $\nu$. Here, it is important to note that the microscope objectives, which we model as aplanatic lenses, do not change the helicity or the AM momentum content of the beam \cite{Ivan2012PRA,Bliokh2011,ZambranaThesis}. That is, using eq.(\ref{S_comm}) notation, $\left[ \Slens, R_z \right]=\left[ \Slens, J_z \right]=\left[ \Slens, \Mz \right]=0$. This is schematically depicted in Fig.\ref{set-up}c, where it can be seen that the incident red helix at the back of the MO$_1$ keeps its color when it is focused by it. Hence, given an incident beam of the kind $\Emp$, the focused beam, denoted as $\Efmp$, will keep the eigenvalues of $J_z$ and $\Lambda$ equal to $m$ and $p$. Then, the transmitted field through a nanohole due to the focused vortex beam can be computed as $\Etfmp=\Sop \left\lbrace \Efmp \right\rbrace$. Here, notice that the subindices $m,p$ refer to the eigenvalues of $J_z$ and $\Lambda$ for the incident beam. Notwithstanding, as mentioned earlier, due to the cylindrical symmetry of the problem, $\Etfmp$ will also be an eigenstate of $J_z$ with value $m$. However, the helicity of the incident beam is not preserved in the interaction. This is a consequence of the fact that duality symmetry is broken by the nanohole and the multilayer system \cite{Ivan2012PRA,Ivan2013,Zambrana2013OE,Bliokh2013,Nora2014}. After the light-matter interaction has taken place, MO$_2$ collects most of $\Etfmp$ (its NA$=0.9$) and collimates it, thus retrieving a paraxial beam. Because MO$_2$ also preserves helicity and AM, the collimated transmitted field $\Etmp$ keeps the same eigenvalue of $J_z$ as $\Emp$, \textit{i.e.} $J_z \Etmp = m \Etmp$. In contrast, $\Lambda \Etmp \neq p \Etmp$, as the sample scatters light in both helicity components because it is not dual-symmetric \cite{Ivan2012PRA,Ivan2013,Zambrana2013OE,Nora2014}. That is, the transmitted collimated field can be decomposed as:
\begin{equation}
\Etmp = \Etmpp + \Etmpm
\end{equation}
where $\Lambda \Etmpp \approx \overline{p} \Etmpp$ and $\Lambda \Etmpm \approx - \overline{p} \Etmpm$ within the paraxial approximation. This is schematically displayed in Fig. \ref{set-up}c, where it is seen that after the sample $S$ an additional blue helix appears. Now, both $\Etmpp$ and $\Etmpm$ can be modelled using an expression similar to that given by eq. (\ref{Epq}): \renewcommand{\arraystretch}{1.5}
\begin{equation}
\begin{array}{ccc}
\Etmpp & = & A_{m,p}(\rho) \ e^{\left( i(m-p)\phi + i k z \right)}  \sphat \\
\Etmpm & = & B_{m,p}(\rho)  \ e^{\left( i(m+p)\phi + i k z \right)} \smphat 
\end{array} \label{Etmpm}
\end{equation} 
where $A_{m,p}(\rho)$ and $B_{m,p}(\rho)$ depend on the NA of MO$_{1}$ and MO$_{2}$. We will denote $\Etmpp$ as the direct component, since it maintains the polarization state $\sphat$. Consequently, the topological charge of $\Etmp$ is still $m-p$. The other orthogonal component is $\Etmpm$, and we will denote it as crossed component. The crossed component has a polarization state $\smphat$ when the incident state is $\sphat$. Due to the cylindrical symmetry, the value of the AM along the $z$ axis must be preserved. Therefore, as it can be seen in eq.(\ref{Etmpm}), when $p$ changes to $-p$, the topological charge of the beam goes to $m+p$. That is, the crossed component $\Etmpm$ is a vortex beam whose optical charge differs in $2p$ units with respect to the incident beam $\Emp$, or with respect to the direct transmitted component $\Etmpp$. This effect was observed by Chimento and co-workers using an incident Gaussian beam and a $20\mu$m circular aperture \cite{Chimento2012}. Similarly, in \cite{Nora2014}, we measured the same phenomenon using LCP Gaussian beams and nanoholes ranging from 100-550 nm. Here, for the first time, we measure the same phenomenon using different incident vortex beams. That is, we record the far-field intensity patterns of vortex beams propagating through nanoholes. In fact, the far-field patterns are measured for the two transmitted helicity components. Notice that these measurements differ from the work previously done in \cite{Zambrana2014Nat} where only the total transmitted intensity power was measured; and they also differ from \cite{Vuong2010}, where the measurements and simulations were done in the near-field. Now, in order to measure the two transmitted helicity components, we use the CCD camera, QW$_2$, and P$_3$ (see Figure \ref{set-up}a). As mentioned earlier, QW$_2$ does the following polarization transformation: \renewcommand{\arraystretch}{1}
\begin{equation}
\begin{array}{ccc}
\spphat & \rightarrow & \xhat \\
\spmhat & \rightarrow & \yhat
\end{array}
\end{equation}
\begin{figure}[htbp]
\centering\includegraphics[width=8cm]{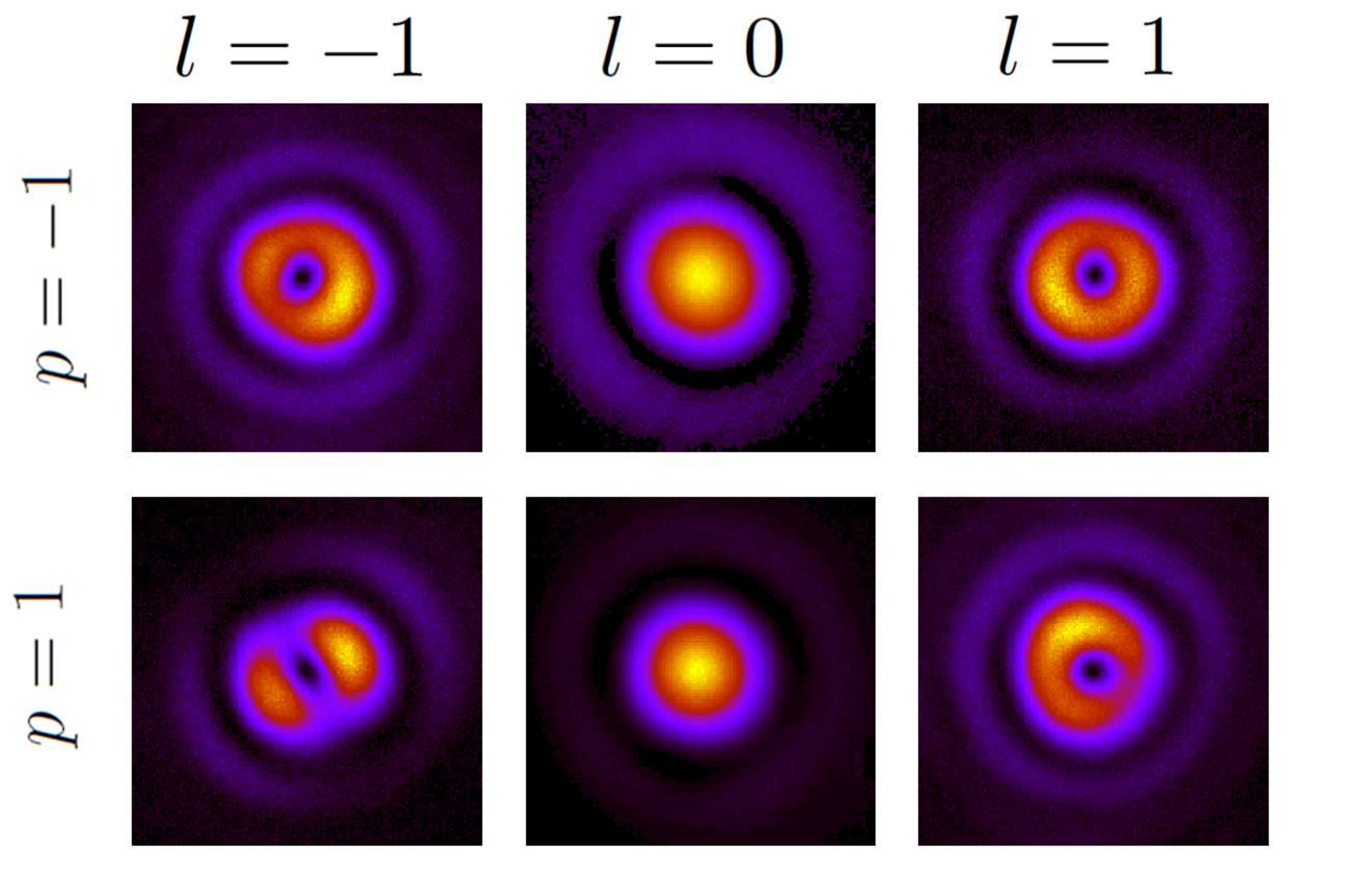}
\caption{Direct component $\vert \Etmpp \vert ^2$ for $p=-1,1$ and $l=m-p=-1,0,1$. The values of $p$ and $l$ shown on the left and the top are the ones carried by the paraxial incident beam $\Emp$. The images have been taken using the nanohole $a_5$ (see Table \ref{Tgammas}). }
\label{Idirect}
\end{figure}
Because both $\Etmpp$ and $\Etmpm$ are paraxial fields, their intensity patterns are not perturbed by this transformation. Thus, their intensity profiles can be singled out by P$_3$. Finally, the CCD camera records the intensity profile of $\Etmpp$ and $\Etmpm$ separately (see Figure \ref{set-up}a). The experiment has been carried out for six different incident fields: three different vortex beams with topological charge $l=m-p=-1,0,1$ are created by the SLM and each of them is right and left circularly polarized ($p=-1,1$). Following the notation in eq. (\ref{Epq}), the six beams used to carry out the experiments are: $\mathbf{E}_{-2,-1}^{\mathbf{in}}, \mathbf{E}_{-1,-1}^{\mathbf{in}},\mathbf{E}_{0,-1}^{\mathbf{in}}, \mathbf{E}_{0,1}^{\mathbf{in}},\mathbf{E}_{1,1}^{\mathbf{in}},\mathbf{E}_{2,1}^{\mathbf{in}}$. In Figure \ref{Idirect}, we show the CCD images of the transmitted direct component $\Etmpp$, when the six incident beams go through $a_5$, \textit{i.e.} a nanohole with diameter $\Phi=333$nm (see Table \ref{Tgammas}). A choice of a different nanohole does not change the images qualitatively. Looking at the intensity patterns displayed in Figure \ref{Idirect}, it can be seen that the direct component of the light transmitted through the nanohole has the same features as the incident beam. That is, both are roughly cylindrically symmetric, they have the same helicity, and the same AM. The AM content of the mode can be inferred from the order of the optical singularity in the center of the beam and the helicity of the mode: following our notation for paraxial vortex beams put forward in eq.(\ref{Epq}), $m=l+p$. In principle, the order of the optical singularity cannot be inferred from an intensity plot. However, in our experiment we are able to confidently assess the absolute value of the order of the singularity. The reason becomes especially clear looking at the images of the transmitted crossed component.
\begin{figure}[htbp]
\centering\includegraphics[width=8cm]{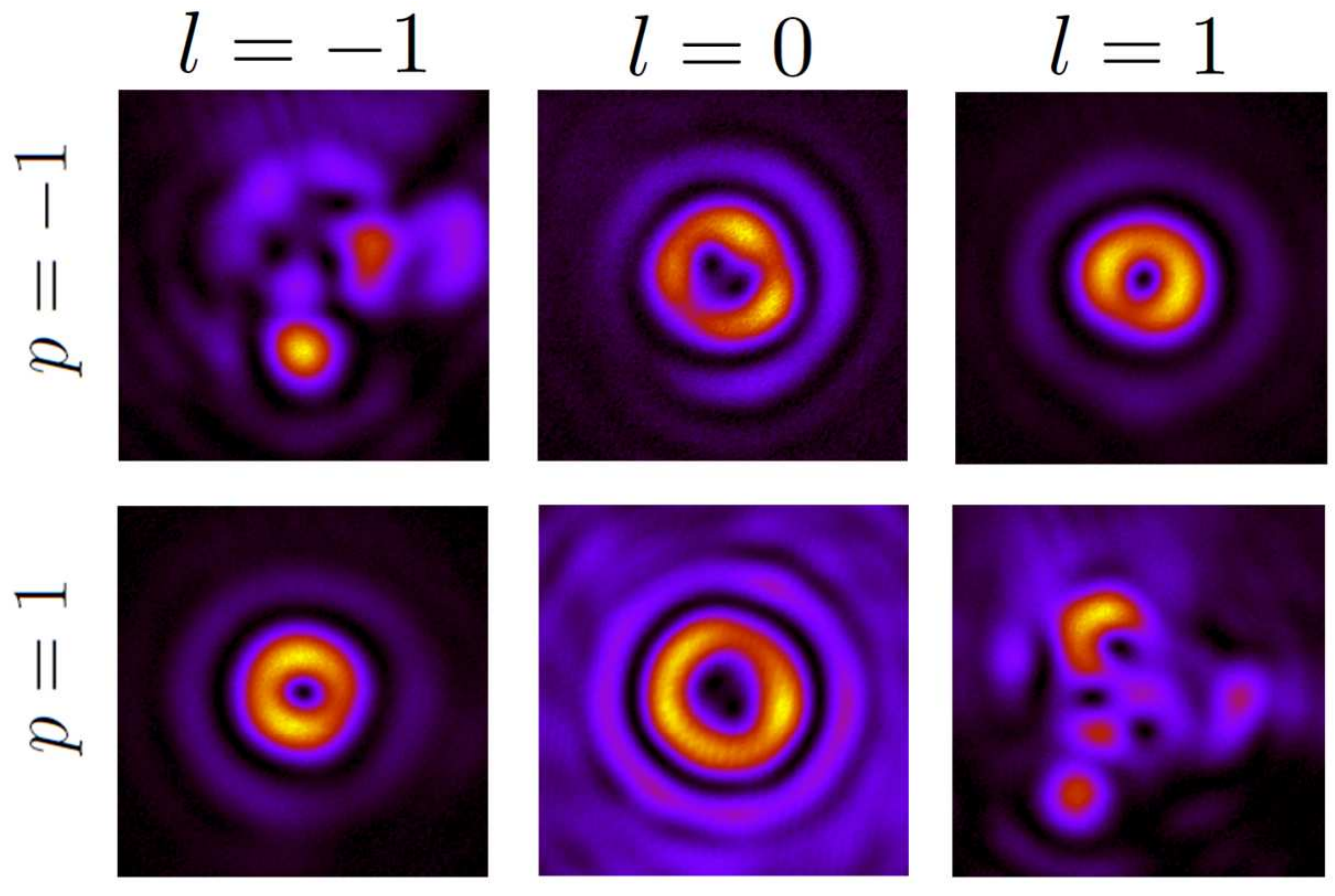}
\caption{Crossed component $\vert \Etmpm \vert ^2$ for values of the paraxial incident field $p=-1,1$ and $l=m-p=-1,0,1$. The values of $p$ and $l$ are the ones carried by the incident beam $\Emp$. Given a row $p$ and a column $m-p$, the image represents $\vert \Etmpm \vert ^2$, which is a mode of light with polarization $-p$ and a phase singularity of order $m+p$. The images have been taken using the nanohole $a_5$ (see Table \ref{Tgammas}).}
\label{I_beam}
\end{figure}
Figure \ref{I_beam} depicts the recorded CCD images for $\Etmpm$ when the values of $p$ and $l$ of the incident beam are $p=-1,1$ and $l=m-p=-1,0,1$. Hence, the image taken  on the row $p=1$ and column $l=1$ corresponds to $\mathbf{E}_{2,1;-1}^{\mathbf{t}}$. That is, $\mathbf{E}_{2,1;-1}^{\mathbf{t}}$ is the crossed component of an incident vortex beam $\mathbf{E}_{2,1}^{\mathbf{in}}$ going through the nanohole. Looking at eqs. (\ref{Etmpm}), it is seen that $\mathbf{E}_{2,1;-1}^{\mathbf{t}}$ should be a vortex beam with a topological charge of order $l'=m+p=3$. However, instead of observing a vortex of charge $l'=3$, three singularities of charge $l=1$ are observed. This occurs because higher order phase singularities are very unstable and prone to split into first order singularities \cite{Gabi2001OL,Ricci2012,Kumar2011,Rich2014}. Thus, in the current scenario, a phase singularity of order $l'$ will split into $\vert l' \vert$ singularities of order $sign(l')$. The instabilities mainly arise from the imperfections in centering the beam with respect to the sample, as well as the tolerances of the linear polarizer P$_3$. Simply put, measuring the number of phase singularities and the helicity of the beam enables us to experimentally verify that the output patterns are consistent with the AM along the $z$ axis being conserved within the experimental errors.

Now, using the method described above, we have consistently measured the intensity ratio between the crossed and direct helicity components. Recently, it has been shown that this ratio is monotonously dependent on the size of the nanohole \cite{Nora2014}. Here, we extend the study presented in \cite{Nora2014}, where the excitation was a LCP Gaussian beam, to different incident vortex beams and we show that the size-dependence is not monotonous. We define this ratio as:
\begin{equation}
\gamma_{m,p} = \dfrac{I_{m,p;\overline{-p}}^t}{I_{m,p;\overline{p}}^t}
\label{gamma}
\end{equation}
where $\Impm$ and $\Imp$ are the intensities of the crossed and direct component measured at the chip of the CCD camera ($A$). Therefore, they can be obtained as: \renewcommand{\arraystretch}{2}
\begin{equation}
\begin{array}{ccc}
\Impm & = & \displaystyle\int_{A} \vert \Etmpm \vert^2 \ {\dd \mathrm{S}} \\
\Imp & = & \displaystyle\int_{A} \vert \Etmpp \vert^2 \ {\dd \mathrm{S}}
\end{array}
\label{inten}
\end{equation} 
Ten single nanoholes have been probed with the same six beams of light used to obtain Figures \ref{Idirect} and \ref{I_beam}.
\renewcommand{\arraystretch}{1}
\begin{table}
\caption{\label{Tgammas}Measurements of $\gamma_{m,p}(\%)$ as a function of the diameter of the nano-aperture for six different incident paraxial beams with $l=m-p=-1,0,1$ and $p=-1,1$. $\gamma_{m,p}$ is computed using eq.(\ref{gamma}).}
\begin{center}
\begin{tabular}{|c|c|c|c|c|}
\hline $\Phi$(nm)& $\sphat$ & $l=1$ & $l=0$ & $l=-1$\\
\hline \multirow{2}{*}{$a_1=212$} & $\spphat$ & $22.0 \pm 0.9$ &  $30.4 \pm 0.5$& $82 \pm 7 $ \\
& $\spmhat$ & $22 \pm 4$ & $29.5 \pm 0.4$ & $15.3 \pm 1.9$ \\
\hline \multirow{2}{*}{$a_2=237$} & $\spphat$ & $22.8 \pm 1.3$ & $22.7\pm 0.4$ & $69 \pm 6$ \\
& $\spmhat$ & $55 \pm 10 $ & $23.5 \pm 0.3$ & $29 \pm 2$ \\
\hline \multirow{2}{*}{$a_3=317$} & $\spphat$ & $23 \pm 2$ & $12.8 \pm 0.2$  & $219 \pm 18 $ \\
& $\spmhat$ & $161 \pm 4$ & $12.3 \pm 0.1$ & $22.5 \pm 1.4$ \\
\hline \multirow{2}{*}{$a_{4}=325$} & $\spphat$ & $18.1 \pm 1.9$ &$14.5 \pm 0.3$ & $360 \pm 30$ \\ 
& $\spmhat$ & $330 \pm 30$ &$14.1 \pm 0.2 $& $22.1 \pm 1.7$ \\
\hline \multirow{2}{*}{$a_5=333$} & $\spphat$ & $39 \pm 3$ & $13.3 \pm 0.1$ & $210 \pm 50$ \\
& $\spmhat$ & $269 \pm 18$ & $12.7 \pm 0.1$ & $49 \pm 8$ \\
\hline \multirow{2}{*}{$a_6=341$} & $\spphat$ & $9.9 \pm 1.3$ & $13.2 \pm 0.1$ & $160 \pm 40$ \\
& $\spmhat$ & $160 \pm 60 $ & $13.7 \pm 0.2$ & $8.0 \pm 1.0$ \\
\hline \multirow{2}{*}{$a_7=424$} & $\spphat$ & $8.4 \pm 0.9$ &  $7.6 \pm 0.2$ & $351 \pm 14$ \\
& $\spmhat$ & $325 \pm 8$ &   $8.2 \pm 0.3$ & $9.7 \pm 0.7$ \\
\hline \multirow{2}{*}{$a_8=429$} & $\spphat$ & $4.7 \pm 0.5$ &  $9.9 \pm 0.2$ & $290 \pm 20$ \\
& $\spmhat$ & $264 \pm 13 $ &  $9.5 \pm 0.2 $& $4.2 \pm 0.8$ \\
\hline \multirow{2}{*}{$a_9=432$} & $\spphat$ & $5.2 \pm 1.4$ &  $8.1 \pm 0.2$ & $289 \pm 9$ \\
& $\spmhat$ & $325 \pm 12$ & $6.9 \pm 0.4$ & $4.6 \pm 0.8$ \\
\hline \multirow{2}{*}{$a_{10}=433$} & $\spphat$ & $8.5 \pm 1.1$ & $7.5 \pm 0.3$& $205 \pm 6$ \\
& $\spmhat$ & $192 \pm 7$ &  $7.9 \pm 0.1$ & $7.9 \pm 0.9$ \\
\hline
\end{tabular}
\end{center}
\end{table}
The results are presented in Table \ref{Tgammas}. The measured $\gamma_{m,p}$ are listed as a function of parameters that we control in the experiment: the topological charge given by the SLM ($l=m-p$) of the paraxial incident beam ($\Emp$) and its circular polarization vector $\sphat$.
\begin{figure}[htbp]
\centering\includegraphics[width=8cm]{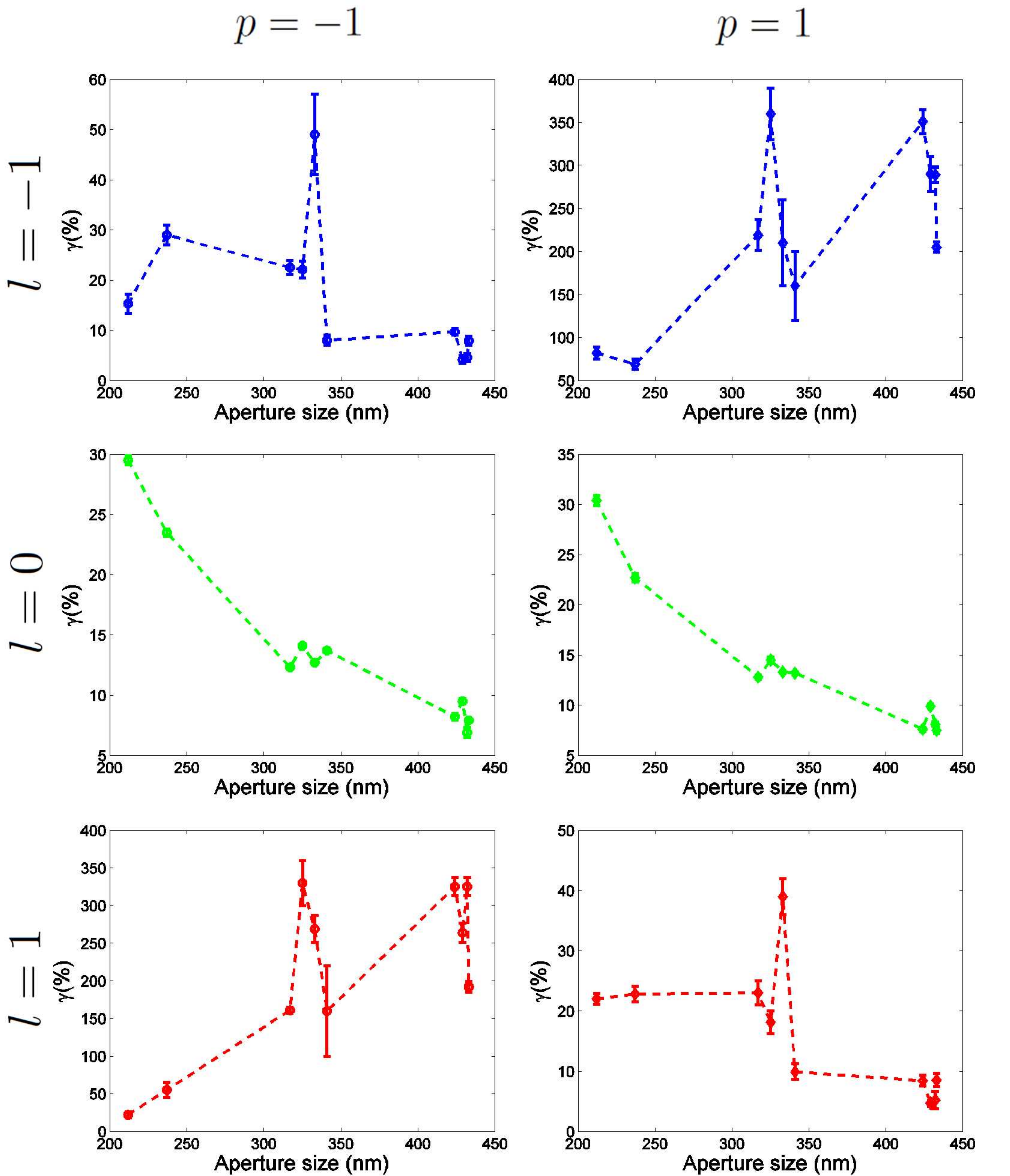}
\caption{$\gmp$ as a function of the diameter $\Phi$. The six different plots correspond to six different paraxial incident beams with $p=-1,1$ and $l=-1,0,1$. The plots from incident beams with the same helicity are listed in the same column. The plot belonging to the same $m-p$ value are displayed in the same row, and plotted in the same color.}
\label{gammas_3c}
\end{figure}

In Figure \ref{gammas_3c} we plot the data from Table \ref{Tgammas}. First of all, it can be observed that the behavior described in \cite{Nora2014} is retrieved for both helicity components when $l=0$. That is, when a circular nano-aperture is excited by a Gaussian beam, its $\gmp$ ratio monotonously decreases from $\gamma_{p,p}=100\%$ for small holes (with respect to the wavelength) to $\gamma_{p,p} \approx 0\%$ for large holes. In fact, it is observed that the result is helicity-independent, as both helicity components yield a very similar result. Nevertheless, the behavior of $\gmp$ for other $\Emp$ modes is more complex. The first striking feature that can be readily observed is that $\gamma_{m,p}$ are no longer monotonous functions of the diameter of the nanohole ($\Phi$). Even though the particular case of a Gaussian beam can be explained looking at the two limits (dipolar behavior for small sizes, and diffraction theory for the large ones) \cite{Nora2014}, it is clear that a more thorough study is needed to explain the behavior of vortex beams through single nanoholes. In particular, these results suggest that vortex beams with $\vert m \vert \neq 1 $ can resonantly couple to a nanohole.

\begin{figure}[htbp]
\centering\includegraphics[width=8cm]{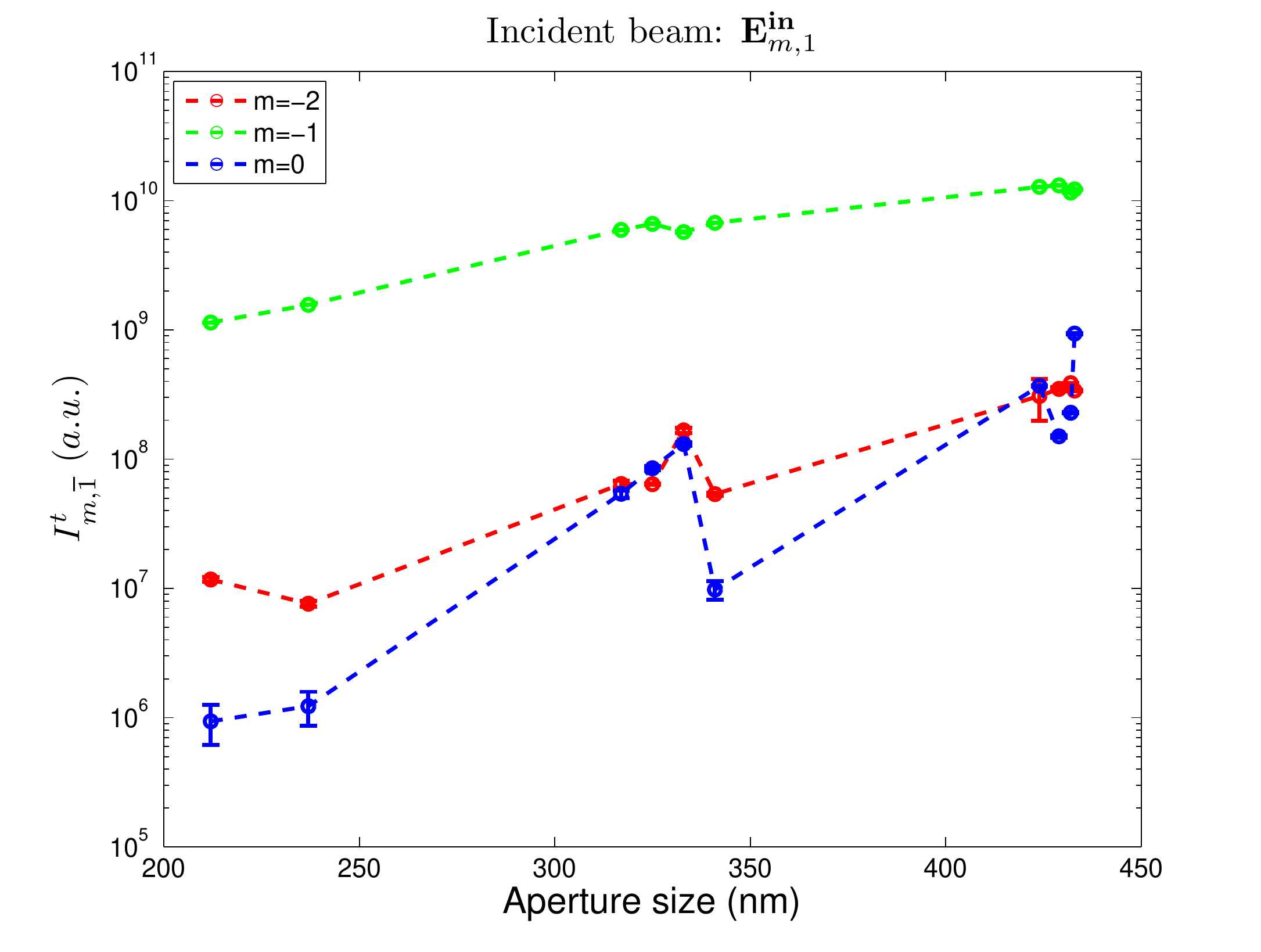}
\caption{$I^{t}_{m,p}$ as a function of the diameter $\Phi$. The three curves are obtained using incident beams with helicity $p=1$, which corresponds to LCP when the beam is paraxial. Each color corresponds to a different $m$ value, as indicated in the inset. The units of the intensity are arbitrary, \textit{i.e.} they are given by the CCD camera. }
\label{LCP}
\end{figure}

\begin{figure}[htbp]
\centering\includegraphics[width=8cm]{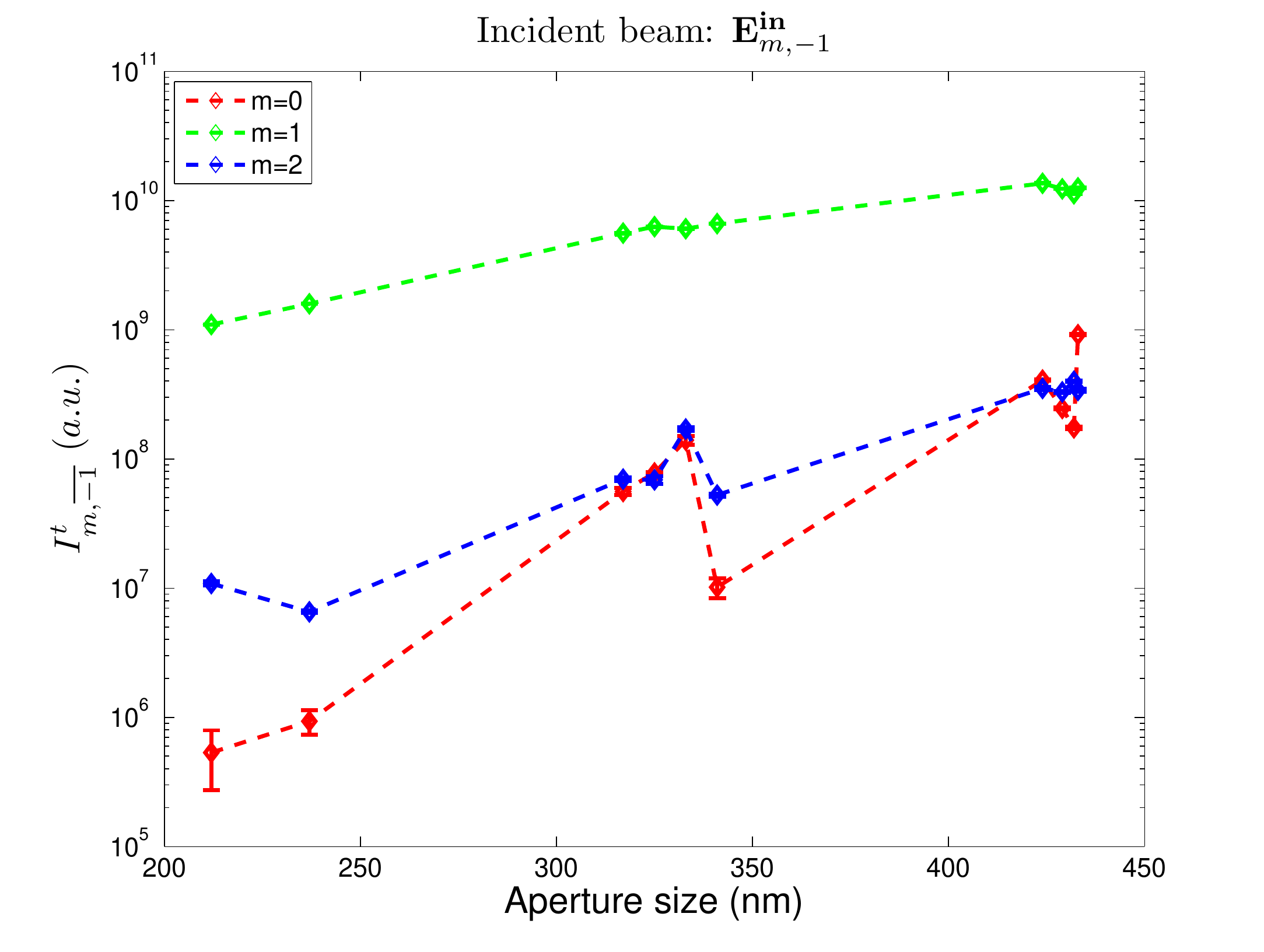}
\caption{$I^{t}_{m,p}$ as a function of the diameter $\Phi$. The three curves are obtained using incident beams with helicity $p=-1$, which corresponds to RCP when the beam is paraxial. Each color corresponds to a different $m$ value, as indicated in the inset. The units of the intensity are arbitrary, \textit{i.e.} they are given by the CCD camera.}
\label{RCP}
\end{figure}

Another interesting feature is that, while in the case of an incident Gaussian beam ($l=m-p=0$), it is seen that the $\gamma_{m,p}$ are independent of the polarization of the incident beam, when the incident field is a vortex beam, this is no longer the case. Indeed, the first and third rows of Figure \ref{gammas_3c} clearly show that given a paraxial vortex beam with charge $l=m-p$, its transmission through a single nanohole strongly depends on the value of the helicity $p$ of the beam. One could expect that in order to study this phenomenon, the structure of the fields at the focus should be studied. However, as it is explained hereafter, the qualitative behavior of the interaction can be understood with only a symmetry discussion. Indeed, the reason behind this different behavior of the propagation of vortex beams through nanoholes is an underlying symmetry relating the transmission of incident beams of the kind $\Emp$ and others of the kind $\Emmmp$. More specifically, both the intensity plots shown in Figures \ref{Idirect} and \ref{I_beam}, as well as the $\gmp$ plots displayed in Figure \ref{gammas_3c} show that the six incident beams can be grouped in three pairs: $( \mathbf{E}_{2,1}^{\mathbf{in}}, \mathbf{E}_{-2,-1}^{\mathbf{in}} )$, $( \mathbf{E}_{1,1}^{\mathbf{in}}, \mathbf{E}_{-1,-1}^{\mathbf{in}} )$, and $( \mathbf{E}_{0,1}^{\mathbf{in}} \mathbf{E}_{0,-1}^{\mathbf{in}} )$, where each pair has a very distinct behavior, whereas both members of each pair share the same features. Note that each member of the pairs shares the same $\vert m \vert$. The formal proof of why this happens can be found in \cite{Zambrana2014Nat}, but the idea is the following one. If we apply a mirror symmetry operator $\Mz$ to $\Emp$, we obtain $\Emmmp$ except for a phase. That is, the two beams of each of the three pairs above are connected with a mirror symmetry. Then, because the sample is symmetric under $\Mz$, mirror symmetric beams will have mirror symmetric scattering patterns, and they will yield equal intensities. This is clearly featured in Figures \ref{LCP}, and \ref{RCP}. Both figures feature the transmitted intensity through the nanoholes for the three incident vortex beams when $p=1$ (Figure \ref{LCP}), and $p=-1$ (Figure \ref{RCP}). The transmitted intensity is denoted as $I_{m,p}^t=\Imp + \Impm$. The colors chosen for each plot are consistent with those used in Figure \ref{gammas_3c}. Because both figures display one of the two members of each of the three pairs of mirror symmetric beams, both figures yield a great resemblance. It is clear now why the transmittance of a Gaussian beam with a well-defined helicity through a nanohole is not dependent on its helicity value $p$: it is the only case where given a vortex beam with a topological charge $l=m-p=0$, the two modes with helicity $-p,p$ are mirror symmetric.

Figures \ref{LCP} and \ref{RCP} also show that the total transmissivity as a function of the diameter of the four beams whose $\vert m\vert \neq 1$ have a non-trivial behavior. That is, their $I^t_{m,p}(\Phi)$ is not monotonic, unlike the Gaussian beam, whose $I^t_{\vert m \vert =1,p}(\Phi)$ was found to be linear in a log-log plot \cite{Nora2014}. However, even though vortex beams can lead to strong helicity-transfer enhancements (Table \ref{Tgammas} and Figure \ref{gammas_3c}), Gaussian beams still yield a much larger transmission through the nanoholes. 

Finally, it is seen that the $\gmp$ ratio can yield results much larger than $100\%$. This is especially clear for the two modes with $m=0$, which yield a $\gamma_{0,p}$ ratio much larger than $100 \%$. That is, even though all the light incident on the nanohole has helicity $p$, most of the transmitted light flips its helicity value, yielding the opposite value $-p$. This is a big difference with respect to the results reported in \cite{Nora2014} using a Gaussian beam, where values of $\gmp \approx 100 \%$ are only reached when the size of the nanohole is very small with respect to the wavelength.

\begin{figure}[htbp]
\centering\includegraphics[width=8cm]{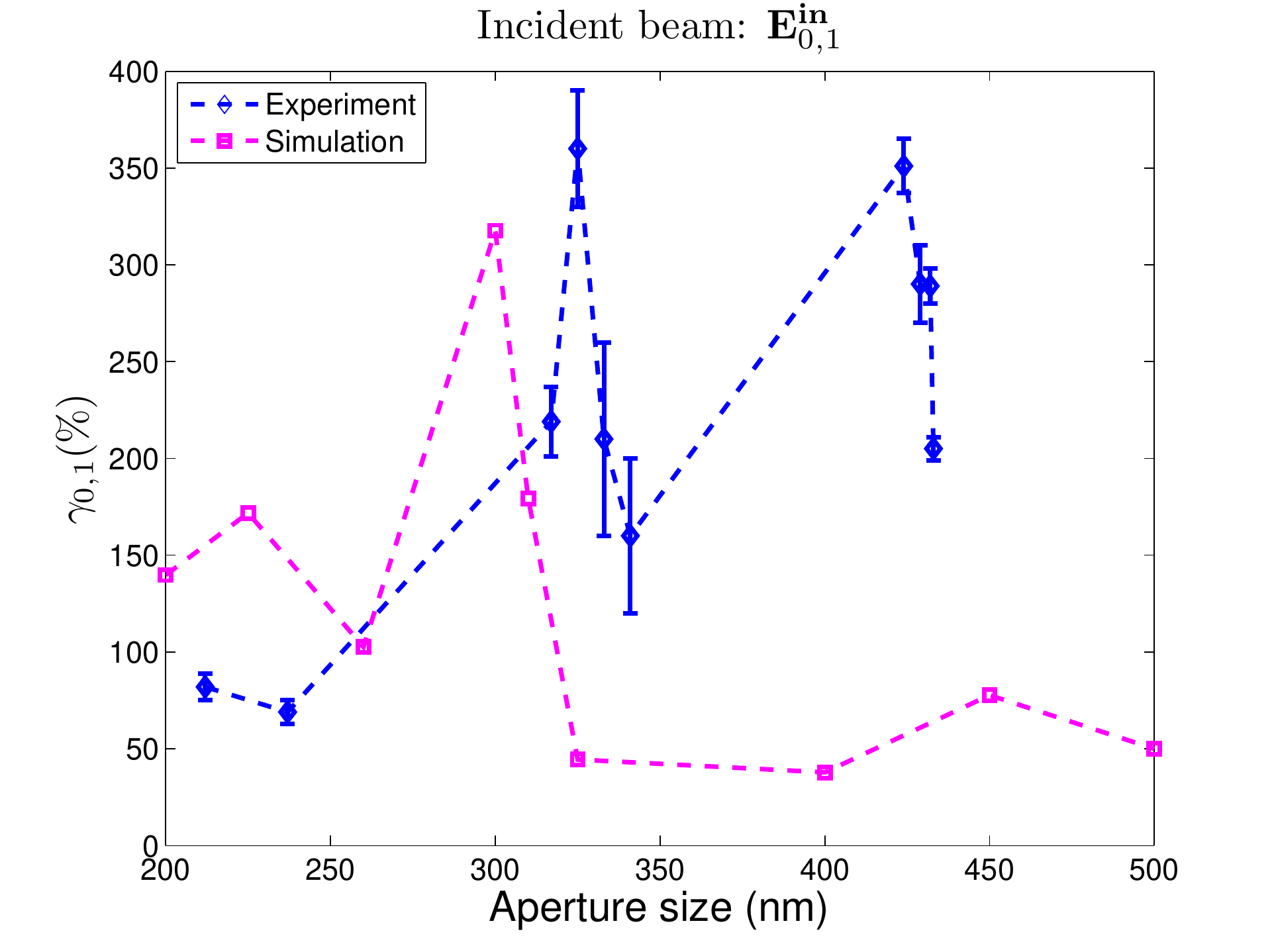}
\caption{$\gamma_{0,1}$ as a function of the diameter $\Phi$. The blue dashed curve corresponds to the experimental values given by Table \ref{Tgammas}. The violet dashed curve has been obtained using the semi-analytical method described in \cite{Ivan2011}.}
\label{resonance}
\end{figure}

Some other systems that have also been reported to produce such large helicity changes are q-plates \cite{Marrucci2006} and dielectric spheres \cite{Zambrana2013OE}. Actually, similarly to the helicity change induced by dielectric spheres, the phenomenon is observed to be very dependent on the size of the structure. That is, as it is depicted in Figures \ref{gammas_3c}-\ref{RCP}, the value of $\gmp$ for the $(\mathbf{E}_{0,1}^{\mathbf{in}}, \mathbf{E}_{0,-1}^{\mathbf{in}})$ increases from about $200 \%$ to $350 \%$ with an increase of $\Delta\Phi \approx 10$nm. This peak in $\gmp$ is very unlikely to be an artefact. Notice that two independent measurements such as $\gamma_{0,1}$ and $\gamma_{0,-1}$ yield analogous results for the same nano-aperture, using completely different incident fields. Furthermore, to verify the strong size-dependence of the helicity transformation in this type of nano-apertures, we have carried out numerical simulations using the semi-analytical method described in \cite{Ivan2011}. The numerical simulations do not exactly reproduced the system described in Figure \ref{set-up}, but they are close enough. The discrepancies in these simulations are mainly two: it has been assumed that the microscope slide was a semi-infinite medium, and the NA of MO$_2$ was NA$=1$. The results of these simulations for the incident beam $\mathbf{E}_{0,1}^{\mathbf{in}}$ are depicted in Figure \ref{resonance}. Even if the peaks do not happen to occur at the same aperture size, Figure \ref{resonance} corroborates that this measurement is not an artefact, but rather an interesting phenomenon to be studied. Nevertheless, a detailed explanation of the fundamental physical mechanism behind this $\gmp$ change deserves a detailed study which is out of the scope of this work.

To conclude, we have shown that vortex beams can go through single nanoholes and have measured their far-field intensity profiles for the first time. We have split their transmitted field into two orthogonal helicity components, and we have observed that the two of them preserve the AM of the initial beam. The ratio between the two transmitted helicity components, $\gmp$, has been measured for six different incident beams. We have observed that their behavior can be grouped into three different sets, each with the same absolute value of the total AM, $\vert m \vert$. It is observed that the sets of beams with $\vert m \vert = 0,2$ have a $\gmp$ behavior very different from $\vert m \vert =1$. Whereas the behavior of the beams with $\vert m \vert =1$ is monotonically decreasing, the other two sets of beams have a non-trivial behavior. In particular, we have measured that the two beams with $m=0$ can yield $\gmp > 300 \%$ for certain sizes of nanoholes. Notice that vortex beams with $m=0$ are already being used as an experimental toolbox for alignment-free quantum communication \cite{Vincenzo2012}. \\

This work was funded by the Australian Research Council Discovery Project DP110103697. G.M.-T. is the recipient of an Australian Research Council Future Fellowship (project number FT110100924).


%

\end{document}